\documentclass[%
reprint,
 amsmath,amssymb,
 aps,
 pra,
floatfix,
]{revtex4-1}

\usepackage{graphicx}
\usepackage{dcolumn}
\usepackage{bm}
\usepackage{hyperref}
\usepackage{aas_macros}
\usepackage{multirow}
\usepackage{comment}
\usepackage{xfrac}
\usepackage[normalem]{ulem}

\usepackage{xcolor}

\def\boxit#1#2#3{%
    \smash{\color{#3}\fboxrule=1pt\relax\fboxsep=2pt\relax%
    \llap{\rlap{\fbox{\phantom{\rule{#1}{#2}}}}~}}\ignorespaces
}

\begin{document}

\preprint{arXiv:xxxx.xxxx}

\title{Astrometry in two-photon interferometry using Earth rotation fringe scan}

\author{Zhi Chen}
 \affiliation{Stony Brook University, Stony Brook, NY 11794}
 \affiliation{Physics Department, Brookhaven National Laboratory, Upton, NY 11973}

\author{Andrei Nomerotski}
\affiliation{Physics Department, Brookhaven National Laboratory, Upton, NY 11973}

\author{An\v{z}e Slosar}
\affiliation{Physics Department, Brookhaven National Laboratory, Upton, NY 11973}

\author{Paul Stankus}
\affiliation{Instrumentation Division, Brookhaven National Laboratory, Upton, NY 11973}

\author{Stephen Vintskevich}
\affiliation{Moscow Institute of Physics and Technology,
Dolgoprudny, Moscow Region 141700, Russia} 

\date{\today}

\begin{abstract}
Optical interferometers may not require a phase-stable optical link between the stations if instead sources of quantum-mechanically entangled pairs could be provided to them, enabling long baselines. We developed a new variation of this idea, proposing that photons from two different astronomical sources could be interfered at two decoupled stations. Interference products can then be calculated in post-processing or requiring only a slow, classical connection between stations. In this work, we investigated practical feasibility of this approach. We developed a Bayesian analysis method for the earth rotation fringe scanning technique and showed that in the limit of high signal-to-noise ratio it reproduced the results from a simple Fisher matrix analysis. We identify candidate stair pairs in the northern hemisphere, where this technique could be applied. With two telescopes with an effective collecting area of $\sim 2$\,m$^2$, we could detect fringing and measure the astrometric separation of the sources at  $\sim 10\,\mu$as precision in a few hours of observations, in agreement with previous estimates.
 \end{abstract}

\maketitle

\section{Introduction}

In a traditional Michelson stellar interferometer \cite{Michelson_interferometer_1, Michelson_inteferometer_2,lawson2000}, a single photon is accepted into two telescope apertures at different locations, and then brought together along two paths to interfere.  The resulting fringe pattern is sensitive to the phase difference incurred due to differences in the photon’s path length to each aperture. 
Assuming that photon occupies a plane wave mode, the interference pattern will be sensitive to changes on an angular scale of $\Delta \theta \sim \lambda/B$ from the source, where $B$ is the baseline, or separation between the apertures, and $\lambda$ is the photon wavelength.  The Michelson setup necessitates maintaining a stable optical path between the stations, which typically limits practical baselines of optical interferometers to a few 100s of meters \cite{tenBrummelaar2005, Pedretti2009, Martinod2018}.

Single-photon optical interferometry is very similar to the radio interferometry, where the wavelength can be of the order of meters to millimeters. In VLBI (Very Long Baseline Interferometry) the baselines can be thousands of kilometers  with observatories spread around the Earth. The astronomical radio VLBI is greatly facilitated by the fact that the electro-magnetic waveforms for the radio frequency scales can be recorded independently and interfered offline later, which is not the case in the optical. Radio VLBI has provided some of the most high-resolution observations in astronomy with one of the recent successes undoubtedly the first image of the supermassive black hole in M87 \cite{akiyama2019first}. However, the resolution in the optical should be better for the same baseline of the same wavelength since it scales with the wavelength. The signal-to-noise ratio cannot be immediately compared between the two, because the emission mechanism, candidate sources and detection techniques are very different in the two cases. Radio VLBI and optical interferometry (both classical single-photon and two-photon version described here) are highly complementary astronomical techniques.


This manuscript discusses a novel type of optical interferometer first proposed in our previous work \cite{Stankus_2022, Nomerotski2020_1} that utilizes quantum  interference effects between two photons from two astronomical sources. By using two sky sources, the proposed interferometer bypasses the traditional necessity of establishing a live optical path connecting detection sites, so the baseline distance can be made, in principle, arbitrarily large, and consequently, an improvement of several orders of magnitude in angular resolution could be attainable.  This development can be considered as a variation of pioneering ideas described in the work of Gottesman, Jennewein and Croke in 2012 \cite{Gottesman2012} to employ a source of path-entangled photon pairs to measure the photon phase difference between the receiving stations, which were further developed in references \cite{harvard1, harvard2, Brown2021, Kwiat2021, Brown2022}. In our case the second sky source is used to produce similar second order correlations of intensity in the two-photon interference.


We note, also, that our technique has considerable overlap with the Hanbury Brown \& Twiss~(HBT) intensity correlation astronomical technique~\cite{hbt, Foellmi2009, dravins2016}, which is used to resolve angular star dimensions by employing two-photon correlation effects. One of the first notable results using the intensity interferometry was to measure the properties of $\alpha$ Virginis binary stars \cite{HerbisonEvans1971} by Herbison-Evan in 1970. 

This technique grew into a mature field over the past decades with applications in numerous projects \cite{abeysekara2020,guerin2018,matthews2018,strekalov2014}. A recent study \cite{bojer2022} also showed that intensity interferometry could provide better astrometry compared to the classical interferometry for close binary star systems, confirming original studies \cite{twiss1969,brown1974, HerbisonEvans1971, hummel1998navy}. In contrast to the classical intensity interferometry, the new approach, advocated here, allows high-precision measurements of relative astrometry for two sources with considerable angular separation. 

In our previous work \cite{Stankus_2022} we proposed to determine the difference in sky positions between the sources employing oscillations of the fringe pattern due to the Earth rotation. This technique of fringe rate measurement allows to improve the astrometric precision as it has more favorable scaling with the experiment duration compared to the standard technique of photon counting.  And since the Earth rotation rate is very stable the measurement can be reliably repeated to study periodic behavior of targeted sources. 


The main purpose of this work is to update a simple Fisher matrix calculation presented in \cite{Stankus_2022} with a more realistic one, and selecting optimal observables for an upcoming on-sky demonstration. Additional goals are also to identify basic parameters of the apparatus that could be able to detect the signal, and to present a concrete algorithm for inferring the parameters of interest from a list of time coincidences of registered photons from the sky sources. For the first time, we aim to determine the applicability of the aforementioned technique employing direct simulations based on the real star pairs and to compare it with theoretical estimations obtained earlier. 

The numerical approach followed here is not meant to model all realistic instrumental effects. It assumes realistic photon fluxes and an actual (simulated) list of events, but the collecting apparatus is still assumed to be ideal. Realistic apparatus effects affect the measurement in at least two ways: (i)~through the total photon counts, which can be reduced due to various inefficiencies such as photon detection inefficiency and transmission losses; and (ii)~spurious background photon detection and atmospheric fluctuation effects \cite{atmospheric_turbulence, atmospheric_scintillation}. The former effect can be incorporated into an effective collecting area and the latter one results in a reduced visibility, so we parametrize and study these effects using these variables.

The rest of the paper is organized as follows: Section \ref{sec:fringe_scan} explains principles of the new two-photon interferometer and Earth rotation fringe scan technique. Then Section \ref{sec:method} describes in detail the observables to be used and their modelling and fitting procedure. Section \ref{sec:selection} discusses selection of possible star pairs for imminent proof-of-principle  observations and numerical simulations. Section \ref{sec:results} describes results of the Markov Chain Monte Carlo (MCMC) procedure \cite{MCMC_overview} which is used to evaluate the astrometric precision for some of the selected pairs. Lastly, Section \ref{sec:conclusions} presents a discussion of the results and conclusions.

\section{Earth rotation fringe rate scanning technique}
\label{sec:fringe_scan}

In this section we briefly review the proposed two-photon interferometer and then describe the Earth rotation fringe scan technique.

\subsection{Two-photon interferometer}

\begin{figure}[h]
\begin{center}
\includegraphics[width=0.9\linewidth]{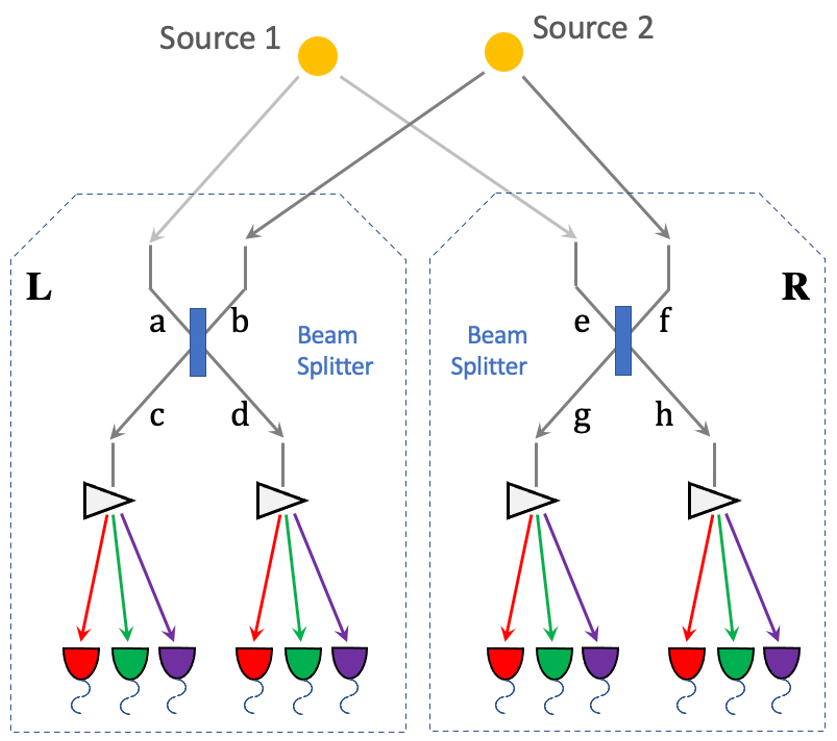}
\caption{Schematic picture of the two-photon amplitude interferometer. We assume that there are two sources which can be observed simultaneously from two stations, \textbf{L} and \textbf{R}, with single spatial mode inputs $a, b$ and $e, f$. Both sources send out photons in the form of plane wave, the path length difference between the two stations yielding phase delays $\delta_{1}$ and $\delta_{2}$ between the photons observed at channels $a,e$ from source 1 and $b,f$ from source 2, respectively. If the two detected photons are close enough in frequency and arrival time, then the pattern of coincidences measured at the outputs $c,d$ and $g,h$ will be sensitive to the difference of the phase differences, $(\delta_{1} - \delta_{2})$, after interference at the symmetric beam splitter in each station.}
\label{fig:idea}
\end{center}
\end{figure}

The basic arrangement of the novel interferometer is shown in Figure~\ref{fig:idea}, following closely the detailed description in \cite{Stankus_2022}: the two sources~1 and~2 are both observed from each of two stations, \textbf{L} and~\textbf{R}.  The key requirement is that photons from Source~1 be coupled into single spatial modes~$a$ at station~\textbf{L} and~$e$ at station~\textbf{R}; while those from Source~2 are separately coupled into the two single spatial modes $b$ and $f$ as shown.
The photon modes~$a$ and~$b$ at station~\textbf{L} are then brought to the inputs of a symmetric beam splitter, with output modes labelled $c$ and $d$; and the same for input modes $e$ and $f$ split onto output modes $g$ and $h$ at station~\textbf{R}.  The four outputs are each viewed by a fast, single-photon sensitive detector.  We imagine that the light in each output port is spectrographically divided into small bins and each spectral bin then constitutes a separate experiment with four detectors.  
If the two photons are close enough together in both time and frequency, then due to quantum mechanical interference the pattern of coincidences between one detector in \textbf{L} (``c'' or ``d'') and one detector  in \textbf{R} (``g'' or ``h'') will be sensitive to the {\em difference} in phase differences $(\delta_{1} - \delta_{2})$ for the two sources; and this in turn will be sensitive to the relative opening angle between them.

\subsection{Fringe rate scanning}

\begin{figure}
\includegraphics[width=1\linewidth]{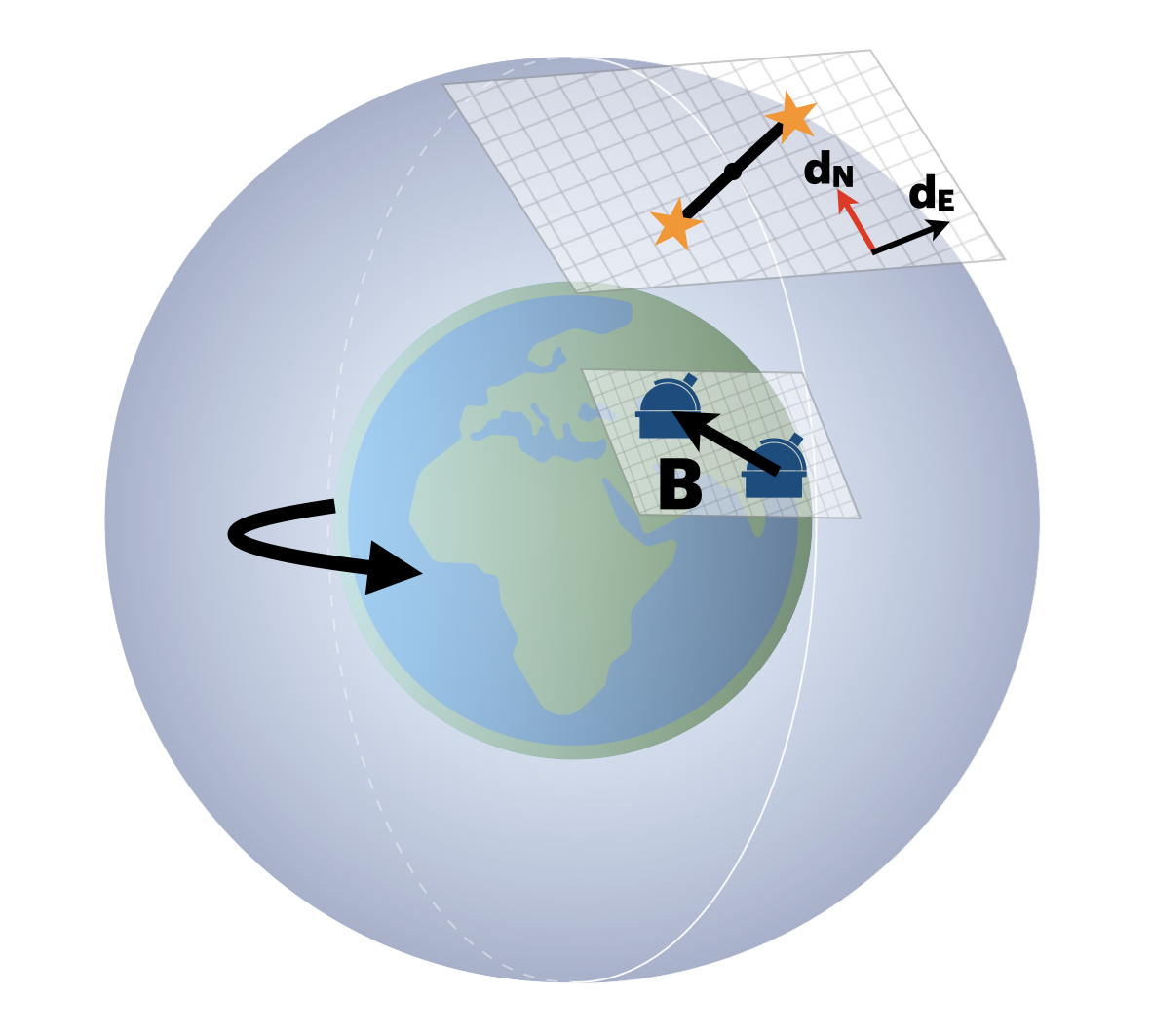}    
\caption{Schematic drawing of the experiment geometry considering an arbitrary baseline and star pair. Two tangent planes are defined by a unit vector point from the center of Earth to the midpoint between the baseline, B, and the separation vector between the two sources. Two coordinates, $d_{N}$ and $d_{E}$, are introduced to characterize the position of the star pair where the origin is defined to be the midpoint of separation vector, which is the black dot between the two stars. $d_{E}$ points in the azimuth direction or into the page, and $d_{N}$ points up to the north pole.
}
\label{fig:coordinate}
\end{figure}

The overall geometry is illustrated in Figure \ref{fig:coordinate}, which shows two stations and two sources on the sky. A tangent plane coordinate system is used to describe the position of the source pair and the baseline of the two telescopes. The origin of the tangent plane is defined to be the mid-point between the two sources. $d_E$ and $d_N$ are the two basis pointing in the azimuth direction and to the north pole, respectively. Let us first consider a simple case that the baseline $B$ between the two stations is straight East-West and both sources lie on the celestial Equator. The light path differences will then be gradually modulated by Earth’s rotation together with observed pair coincidences as a function of time.
As shown in \cite{Stankus_2022}, this fringe angular rate $\omega_{f}$ provides a direct measure of the opening angle between the sources $\Delta\theta$ if all the other parameters are known.
In the limit of small opening angle the fringe rate is just proportional to $\Delta\theta$ 

\begin{equation}
\omega_{f} = \frac{2 \pi B \Omega \sin \theta_{0}  }{\lambda} \Delta\theta ,
\end{equation}
where $\lambda$ is the photon wavelength, $\theta_{0}$ is the position of source 1 at the epoch chosen as $t=0$, and 
$\Omega$=7.29 $\times$10$^{-5}$ rad/sec is the angular velocity of the Earth's rotation.

Generally, measurements of frequency across a time domain are among the most precise allowing us to outline possible strategy of astrometric measurements.  We can make a measurement of $\omega_{f}$ every day at the same $\theta_{0}$; and then day-by-day changes in $\omega_{f}$ over a season would provide information on the evolution of $\Delta \theta$ due to parallax, orbital motions, gravitational lensing, etc., as well as relative overall proper motion. 
\section{Method}
\label{sec:method}

In this section we provide exact expressions for the geometry and explain the measurement methodology including detailed simulations and derivation of the likelihood function.

\subsection{Geometry}
In a single source interferometry, the fringe is cased by changing of the the path length difference in arrival times of the photon to the telescope.  In two source interferometry, we are sensitive to the difference in the two path length differences. In \cite{Stankus_2022} we have considered a simple geometry where both the telescope baseline and the stair pairs are aligned in the East-West direction. We will now calculate this geometry in a general case so we can apply it to the real star pairs. We will work in small angle approximation for both the telescopes (i.e. baseline is much smaller than the radius of the earth) and sources (the inter-source separation in radians is much smaller unity), which should be acceptable for all practical purposes.

Without loss of generality, we can put telescopes at longitude 0 and latitude $\theta_L$. It can be shown that the baseline vector is given by
\begin{equation}\label{Eq:baseline}
    \Vec{B} = B_N \begin{pmatrix}
        -\sin{\theta_L} \\
        0 \\
        \cos{\theta_L} 
    \end{pmatrix}
    + B_E \begin{pmatrix}
        0 \\
        1 \\
        0
    \end{pmatrix}
    + B_V \begin{pmatrix}
        \cos{\theta_L}\\
        0 \\
        \sin{\theta_L}
    \end{pmatrix};
\end{equation}
where $B_N$ and $B_E$ are the North-South and East-West components of the baseline, and $B_V$ is the vertical component of the baseline. We note that $B_V$ is negligibly small for the typical baselines ($\sim 200$ m) that we consider here.

A vector to the pair of source points towards
\begin{equation}\label{Eq:unit_vector}
    \hat{s} = 
        \begin{pmatrix}
    \cos{\delta}\cos{\phi} \\
    \cos{\delta}\sin{\phi} \\
    \sin{\delta}
    \end{pmatrix};
\end{equation}
where $\delta$ is declination of the source pair and $\phi$ is the local hour angle. In this setup, when $\phi=0$, the source transits. We work in the Earth fixed frame, so $\phi$ varies with time as $\phi = \Omega t$, where $\Omega = 2\pi/(1\, {\rm day})$ is the Earth angular velocity. 
By taking derivatives of $\hat{s}$ with respect to $\delta$ and $\phi$ and appropriately renormalizing, we find that the vector connecting the two source can be written as
\begin{equation}\label{Eq:source_vector}
    \Delta \Vec{s} = d_N \begin{pmatrix}
        -\sin{\delta}\cos{\phi} \\
        -\sin{\delta}\sin{\phi} \\
        \cos{\delta} 
    \end{pmatrix}
    + d_E \begin{pmatrix}
        -\sin{\phi} \\
        \cos{\phi} \\
        0 
    \end{pmatrix};
\end{equation}
where $d_N$ and $d_E$ and the source separations in radians towards the North and East on the celestial sphere. 

We thus find that the difference in path difference can be expressed as

\begin{equation}\label{eq:dot_product}
\begin{aligned}
\footnotesize
    \Delta L = 
    \begin{bmatrix}
    B_N \\
    B_E
    \end{bmatrix}^T
    \begin{bmatrix}
    \sin \theta_L \sin \delta \cos \phi + \cos \theta_L \cos \delta & \sin \theta_L \sin \phi \\
    -\sin \delta \sin \phi  & \cos \phi\\
    \end{bmatrix}
    \begin{bmatrix}
    d_N \\
    d_E
    \end{bmatrix}
\end{aligned}
\end{equation}

\subsection{Coincidence rate}
\newcommand{\nbar}{{\bar{n}}}
In the two-photon interferometry setup shown in Figure~\ref{fig:idea} there are four types of observed coincident pairs with one photon at each station {\bf L} and {\bf R}, namely $cg$, $dh$, $ch$ and $dg$, referring to the beam splitter output channels in the figure.  The rate for each pair type will show fringes, with $cg$ and $dh$ both moving together and $ch$ and $dg$ both moving oppositely:

\begin{equation}
    R_{\pm}(t)  = \bar{n} \left ( 1 \pm V \cos{\left(\frac{2\pi \Delta L}{\lambda} + \psi\right)}\right), \label{eq:rate}
\end{equation}
where, in notation of \cite{Stankus_2022}:
\begin{eqnarray}
R_+ &=& n_{cg}+n_{dh} \nonumber \\ 
R_- &=& n_{ch}+n_{dg} \label{eq:twoset},
\end{eqnarray}
and $V$ is the fringe visibility (see below).
Here we use $n_{xy}$ for the rate of pair coincidences in the $x$ and $y$ outputs, $R_{\pm}$ is the combined rate for each co-moving set of pairs, and $\bar{n}$ is the fringe-averaged value of $R$.
 \footnote{Note that $\nbar$ has different normalization in this paper compared to \cite{Stankus_2022}}

We can then relate $\bar{n}$ to the source intensities and the assumed parameters of the light collection and single photon detection: 
\begin{equation}
    \nbar = \frac{(S_1+S_2)^2\tau}{4} \left(\frac{A \Delta \nu}{2 h \nu}\right)^2 ,
\end{equation}
where $S_{1,2}$ are the fluxes of corresponding sources expressed as energy per unit time per unit collecting area and per unit bandwidth; $A$ is the effective collecting area at each of the four apertures, here assumed to be identical; $\nu$ and $\Delta \nu$ are the photon frequency and allowed bandwidth; and $\tau$ is the arrival time window over which the two photons can be considered in coincidence.  

To observe quantum interference between photons, path alternatives for the photons in output modes after the beamsplitters must be indistinguishable, meaning that the product of the differences
in their measured arrival times $\Delta t$ and their frequencies $\Delta \nu$ must satisfy $2 \pi \Delta t \, \Delta \nu  < \, \sim 1 $.
This means that if our detector time resolution is $\tau$ then the natural choice of frequency bin width $\Delta \nu$ will be $\Delta \nu = (2\pi \tau)^{-1}$, which is what we use in the simulations here. The total signal grows linearly with bandwidth. The factor of 8 stems from considering the number of all possible quantum-mechanically allowed photon coincidences. The factor in brackets converts the flux in Jy to the number of photons per unit time.  The factor of 2 in that denominator accounts for the fact that conventionally $S$ is the the total flux density over both photon polarizations, while the interference condition requires both photons to be in the same polarization mode, and so we assume only one polarization is accepted.

Classically, $V = 2 S_1 S_2 / (S_1+S_2)^2$, which is maximized at $V=\sfrac{1}{2}$ when $S_1=S_2$. A more careful quantum-mechanical calculation for the case of two thermal sources results in $V = 2 S_1 S_2 / ((S_1+S_2)^2+S_1^2+S_2^2)$ in the limit of both sources individually unresolved by the interferometric baseline. This final expression gives a maximum visibility of $V=\sfrac{1}{3}$, but for each case considered in this paper, we evaluated the visibility individually in the fit providing corresponding values. 

\subsection{Simulations}

\begin{figure}
\begin{center}
\includegraphics[width=8.6cm]{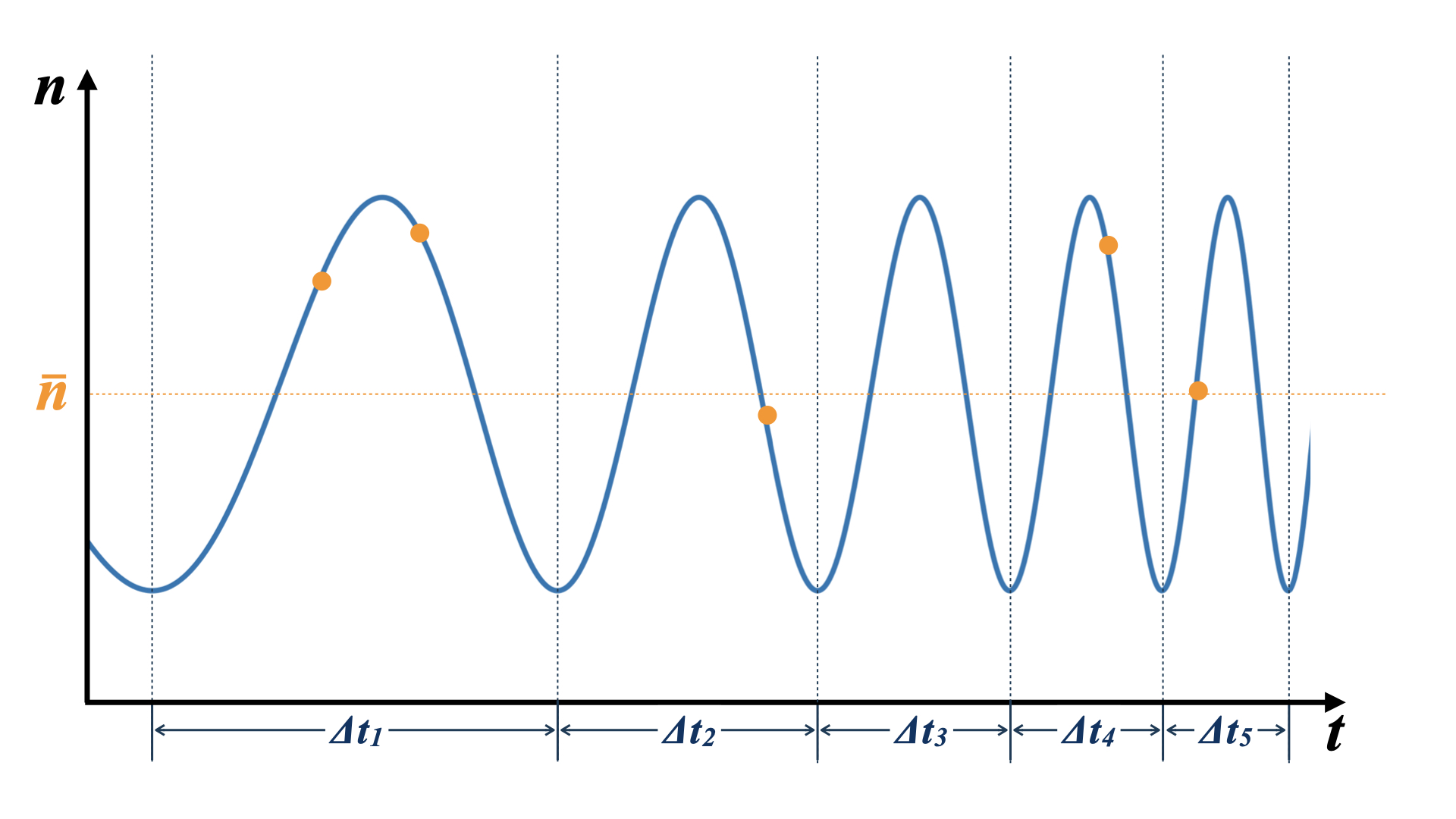}
\caption{Schematic explanation on how simulated coincidences are generated. \emph{i)} We first determine the period of each fringe cycle denoted by $\Delta t$'s, and then the number of event occurrences in each fringe cycle (denoted by the orange points) is determined by Poisson distribution with expected number of coincidences equal to $\nbar \Delta t$. Note that in practice there will be at most one event for each fringe cycle for a realistic $\nbar$ of a typical star pair; \emph{ii)} Next we determine the phase of each event in the cycle through inverse transform sampling from Equation \ref{eq:rate};  \emph{iii)} Finally, we calculate the corresponding timestamp of each event based on their phase.}
\label{fig:simulated_coincidences}
\end{center}
\end{figure}

In order to get datasets as realistic as possible, we need to simulate coincidences that follow the prescribed Poisson process. The fringing frequency that modulates the coincidence rate is considerably faster than the total observation time, which makes brute-force simulation in terms of small time bins impractical. Instead we split the simulation process into two steps: i) identifying which fringe cycles will contain an event and ii) placing events into those cycles. See Figure \ref{fig:simulated_coincidences} for a visual guide on how coincidences are simulated.


We first determine the edge of fringe cycles, i.e. times for which $2\pi \Delta L / \lambda + \psi = 2N\pi$ for an integer $N$,  for a given $\psi$ and $\phi = \Omega t$ (see Eq. \ref{eq:dot_product}). We then calculate the number of event occurrences in each fringe cycle of the coincidence rate curve by sampling from $\nbar \Delta t$, where $\Delta t$ is the duration of that particular cycle. This assumes that the timescale of a single fringe cycle is small compared to the timescale of the Earth rotation, i.e. $\Delta t << 2\pi/\Omega$, where $\Omega$ is the angular velocity of Earth rotation.

The second part involves placing events within each cycle. The relative probability within cycle follows $1\pm V\cos(\alpha)$, where $\alpha=2\pi t/ \Delta t$ measures the phase within cycle. We place an event through inverse transform sampling. A random number r is drawn from a uniform distribution on the interval $[0,1]$ and then $\alpha$ is found so that it solves
\begin{equation}\label{Equation:CDF}
    {\rm CDF}(\alpha) = \frac{\alpha \mp V\sin{(\alpha)} + \pi}{2\pi} = r \quad \quad \quad  \alpha\in[-\pi, \pi]
\end{equation}
In practice, this equation is solved by constructing a set of  cumulative distribution functions ${\rm CDF}(\alpha)$ \cite{miller2010inverse} and interpolating it with an appropriate spline.
The value of $\alpha$ is then converted to the actual timestamp.  While this assumes that $\Delta L \propto \Delta T$ during an individual cycle, which is an excellent approximation, we correctly account for changing cycle duration due to projection effects. 

The result of this process is a collection of time tags for simulated coincidence events for both $R_\pm$ branches of events. We now use these data for constructing a likelihood for a given model.

\subsection{Likelihood}
Coincidences as a Poisson process have a rate, which varies with time. It can be shown that the log probability of obtaining a set of events at times $t_i$ from a Poisson process observed between $t=0$ and $t=T$ is given by 
\begin{equation}
\log P(t_i | R(t)) = \sum_i \log R(t_i) - \int_0^T R(t) dt
\end{equation}
This can be derived by considering a set of finite bins in time and letting the limit of bin width going to zero. Assuming a constant rate $R(t)$, the above expression is maximized for $R=C/T$, where $C$ is the total number of events observed in time T as expected. However, we are not interested in the absolute rates, but instead in the shape of $R(t)$ and therefore we can safely ignore the second term.

 In our case, R depends on 4 parameters: $\theta = \{ d_N, d_E, \psi, V\}$. The Bayes theorem for the probability of these events requires the log likelihood 
\begin{equation}
    P(\theta | t^+, t^-) \propto (t^+, t^- | \theta) = \sum_i \log R_+(t^+_i) + \sum_i \log R_-(t^-_i), \label{eq:loglike}
\end{equation}
where $t^\pm_i$ are the events occurring the in $\pm$ branches of data,
Fortunately, at the level of sensitivity under consideration, the Equation \ref{eq:loglike} can be evaluated numerically for a reasonable number of events. This is a very important aspect, because it allows us to work directly with the likelihood, rather then with phenomenological algorithms. This ensures optimality of the analysis.

\subsection{Posterior exploration}
We employ the likelihood derived above in a Markov-Chain Monte Carlo (MCMC) procedure \cite{10.2307/2242477}. We use a simple Metropolis-Hasting algorithm as implemented in the \texttt{April} software packages \cite{April}. In total, there are four parameters to vary and perform the error estimation in the MCMC simulation. The two-point source visibility, or $V$, is the first parameter which contains the information of the relative brightness between the two sources. There are two parameters, $d_{N}$ and $d_{E}$, corresponding to relative separation between the sources. Since $d_{N}$ and $d_{E}$ are the separation of North-South and East-West direction of the tangent plane, it is necessary to provide the initial midpoint position, $\delta_{mid}$ and $\phi_{mid}$, between the two sources for the MCMC simulation to work. The last parameter is a constant offset phase, $\psi$, which is assumed to be unknown and corresponds to an unaccounted residual path length between the two stations.  The likelihood is calculated using Equation \ref{eq:loglike}. After running the MCMC procedure, we plot the posterior estimates using the \texttt{corner} package and find  marginalized errors of different parameters by calculating the square root of the corresponding diagonal elements of the covariance matrix.

\begin{table*}[t] 
\caption{\label{tab:BSC}%
This table shows the ideal star pair for real-world astrometry through two-photon interferometry depending on time of the year. The HR \# column refers to the Harvard Revised Number (Bright Star Number) of each star. The corresponding right ascension, declination, and visual magnitude of each star pair are listed. The relative separation between each pair is described using the tangent plane system, where $d_{N}$ and $d_{E}$ are the North-South and East-West separation each pair in the tangent plane. Note that the upper value of $d_{E}$ and $d_{N}$  corresponds to $d_{E}$ and the lower value corresponds to $d_{N}$. Note that due to finite number of digits in the catalog, these are accurate to about 0.2 arcsec. The spectral flux density of each star is calculated using Equation \ref{Eq:mag_to_flux} assuming a visual band filter with $F_{V,0} = 3640$ Jy. The red and blue box indicate the 15 arcsec and 1 arcsec star pair that we used for simulation. The last column shows the typical $V/\sigma_V$ ratio calculated from the MCMC simulation for a typical setup with fixed effective collecting area times number of channels of 3.5\,m$^2$.
}

\begin{ruledtabular}
\centering
\begin{tabular}{cccccccc}

Optimal& HR \# & Right Ascension & Declination & Flux Density & $d_{E}$ and $d_{N}$ & Visual Magnitude & $V/\sigma_V$ \\
Visibility &  & [rad] & [rad] & [Jy] & [arcsec] &  &\\

\colrule
\multirow{2}{*}{Jan}     & 4058 & 2.70516 & 0.34628 & 109.93 & $4.2$ & 3.80 & \multirow{2}{*}{3.08}\\
                         & 4057 & 2.70513 & 0.34630 & 328.93 & $-4.0$& 2.61 &                  \\
                         \hline

                         & 5055 & 3.50785 & 0.95856 & 95.74 & $7.8$ & 3.95 & \multirow{2}{*}{2.59}\\
\boxit{6.7inch}{0.195inch}{red} Feb-Mar  & 5054 & 3.50778 &  0.95863 & 449.88 & $-13.0$ &  2.27 & \\
                         \hline

\multirow{2}{*}{Apr}     & 5506 & 3.86148 & 0.47253 & 302.76 & 0.0 & 2.70 & \multirow{2}{*}{0.92}\\
                         & 5505 & 3.86148 & 0.47255 & 32.59  & $-3.0$ & 5.12 &\\
                         \hline

\multirow{2}{*}{May-July}  & 7956 & 5.44185 & 0.59994 & 39.18 & $720.9$ & 4.92 & \multirow{2}{*}{1.29}\\
                         & 7949 & 5.43762 & 0.59289 & 377.66 & $1454.0$ & 2.46 &\\
                         \hline

\multirow{2}{*}{Aug-Oct} & 604  & 0.54066 & 0.73881 & 42.18 & $7.8$ & 4.84 & \multirow{2}{*}{1.42}\\
                         & 603  & 0.54062 & 0.73879 & 454.04 & $4.0$ & 2.26 &\\
                         \hline

                         & 2891 & 1.98357 & 0.55655 & 587.63 & 0.0 & 1.98 & \multirow{2}{*}{7.16}\\
\boxit{6.7inch}{0.195inch}{blue} Nov-Dec  & 2890 & 1.98357 & 0.55656 & 256.51 & $-1.0$ & 2.88 & \\
                         
\end{tabular}
\end{ruledtabular}
\end{table*}

\section{Selection of observable star pairs}
\label{sec:selection}

We investigated potential targets for a real-world astrometry pathfinder for the proposed two-photon interferometer. Our source catalog is the Yale Bright Star Catalog \cite{BSC}, which lists over 9,000 stars with visual magnitude of 6.5 or brighter together with various properties of each individual star such as the right ascension, declination, galactic latitude, galactic longitude, visual magnitude, rotational velocity, radial velocity etc. 

For a concrete example of a pathfinder experiment on continental US, we search for the following pairs of stars:
\begin{itemize}
    \item Pair needs to be in the northern hemisphere due to geographical reasons;
    \item Separation of stars in the pair is less than 0.01 radians ($\sim 0.6$ degrees). This ensures that both stars should fit comfortably in field of view of a typical high-end amateur grade telescope;
    \item Pair separation of more than 1 arcsec that should enable a relatively clean separation of light under good atmospheric conditions. 
\end{itemize}

Only the stars that are 12 to 17 hours behind the sun are considered in order to reduce unwanted background from the Sun. Since the relative angular separation between the Sun and stars gradually changes over the year, we make a rough estimation that the delay time of each star is decreased by 2 hours for every month away from the vernal equinox. Finally, the star pair that has the largest average spectral flux density is selected to maximize the expected coincidence photon rates.

Table \ref{tab:BSC} shows the star pairs, which are optimal for observations for the corresponding month. It also shows different properties of each pair including the right ascension, declination, and visual magnitude. The relative separation of each star pair described in the tangent plane coordinate system is also shown in Table \ref{tab:BSC}, where $d_{N}$ and $d_{E}$ are the separation in the North-South and East-West direction of the tangent plane which is defined by an unit vector pointing to the midpoint between the two sources. The flux density of each individual star is converted from visual magnitude to flux density using Equation \ref{Eq:mag_to_flux} where $F_{x,0}$ is the reference flux of different band filters at zero apparent magnitude. 

\begin{equation}\label{Eq:mag_to_flux}
    F = F_{x,0} \times 10^{m/-2.5}
\end{equation}

Table \ref{tab:BSC} shows the corresponding spectral flux density of each star assuming a visible band filter of 0.55 $\mu$m, which has a corresponding $F_{\textrm{visible},0} = 3640$ Jy. The red and blue boxes indicate the 15 arcsec and 1 arcsec star pairs which are the example star pairs, which we investigated further in Section \ref{sec:realistic_cases}. The last column shows the typical visibility to $\sigma_V$ ratio calculated from the MCMC simulation using the following setup: an East-West baseline of 200 m, $A$ or effective collecting area of 3.5 $m^2$ with 100 \% efficiency, $\tau$ or time bin of correlation of 0.15 ns, $\Delta \nu$ or detector bandwidth of 1 GHz, $\lambda$ or wavelength of observation of 0.55 $\mu$m, latitude of the observatory at 40.7 degree North, and a total observation period of 20,000 s.

\section{Results}
\label{sec:results}

\subsection{Comparison to Fisher matrix predictions}
We start by cross-checking our results against Fisher matrix predictions of the signal-to-noise errors presented in our previous paper \cite{Stankus_2022}. There we assumed that the baseline is purely East-West, the observatory is located on the equator and both sources have zero declination. The resulting expression gives error as 
\begin{equation}
\begin{gathered}\label{Eq:fisher_error}
    \sigma[\Delta \theta] = \sqrt{\frac{6}{\pi^2 \kappa}} \frac{\Delta \theta}{V N^{Cycle} \sqrt{2\bar{n}T}} \\
    \kappa[V] = \frac{1-\sqrt{1-V^2}}{V^2},
\end{gathered}
\end{equation}
where $T$ is the observation time, $N^{\rm Cycle}$ is the number of fringe cycles and the other quantities are as defined before. 
The usual caveats associated with the Fisher matrix formalism apply. In particular, it is known that the Fisher matrix error estimates are only valid when the signal-to-noise is sufficient to the extent that the posterior probabilities are Gaussians and the quadratic expansion of likelihood around the fiducial models correctly reflects the uncertainties. Therefore, we expect the Fisher and full likelihood analyses to agree at high signal-to-noise ratio with some additional differences stemming from approximations such as the fixed fringe rate for the former. Moreover, we also need to account for an extra factor of $1/\sqrt{2}$ compared to what we had previously because now we are considering both $\pm$ branches of the data as described by Equation \ref{eq:loglike}. 

\begin{figure}

\includegraphics[width=1\linewidth]{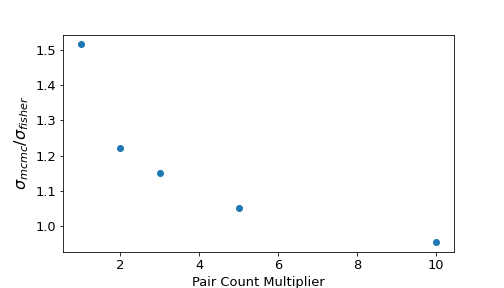}
\caption{Comparison between the error estimation from the MCMC simulation to the error estimation using the Fisher's analysis from our previous work \cite{Stankus_2022} for different pair count multiplier. The vertical axis shows the ratio between the separation error of the two sources in the East-West direction from calculating the square root of the covariance matrix of the MCMC simulation and using Equation \ref{Eq:fisher_error}. The horizontal axis shows the coincidence pair rate multiplier used for boosting the sample data size. As the coincidence pair multiplier increases, the ratio gradually approaches 1 as expected.}
\label{fig:fisher_crosscheck} 
\end{figure}

We looked at the same setup that we mentioned in Section \ref{sec:selection}, except $A$ or effective collecting area of 2.8 $m^2$ with 100 \% efficiency and assume the observatory is located on the Equator, i.e. latitude of 0 deg. We picked the star pair HR \# 5054 and 5055 from the bright star catalog (indicated by the red box in Table \ref{tab:BSC}) and manually changed their declination to 0. Using these settings, we tested the validity of error estimation of MCMC simulation by manually increasing the coincidence pair rate of the two sources by different factors and compared the errors in the estimated uncertainties. Results are plotted in the Figure  \ref{fig:fisher_crosscheck}. We observed that our MCMC formalism reproduces the Fisher matrix result to better than 5\% for high signal-to-noise ratio, which gives confidence in both the MCMC and Fisher matrix calculations. As expected, the Fisher matrix fails to encompass the full uncertainty at lower signal-to-noise because it does not correctly account for long tails associated with borderline detection.

\begin{figure*}
\centering
\includegraphics[width=8.6cm]{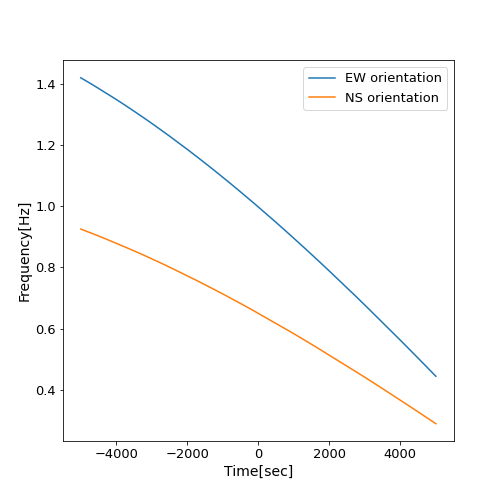}
\includegraphics[width=8.6cm]{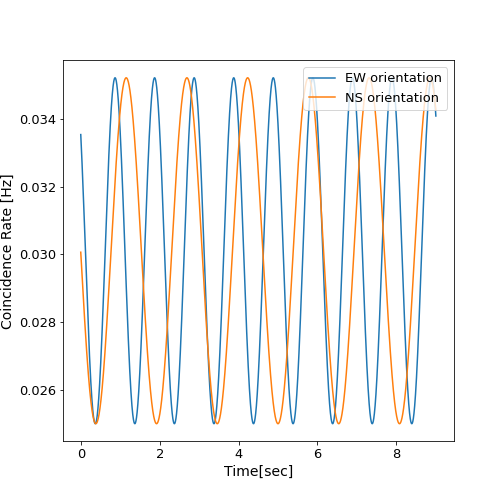}
\caption{
{\emph{Right}: Theoretical fringe patterns for the 15 arcsec separation star pair with East-West and North-South baseline orientation. \emph{Left}: Frequency of the signal changes over time. We observe that the oscillation frequency is higher for the East-West baseline compared to the North-South baseline as expected.}}
\label{fig:ew_vs_ns} 
\end{figure*}

\subsection{Realistic Cases from Bright Star Catalog}\label{sec:realistic_cases}

In this section, we explore the minimum effective collecting area of the telescope with 100 \% efficiency to generate well-constrained triangle plots to characterize the distribution and correlation of parameters using two real star pair cases selected from the bright star catalog that are separated by 1 arcsec and 15 arcsec. These two pairs are the ideal star pairs for observation from February to March and November to December, respectively, as shown in Table \ref{tab:BSC}. In this simulation, the same parameter settings are used in the previous section except now we vary the collecting area, assuming the observatory is located at the New York latitude of 40.7 deg North, and doing cases for the baseline of 200 m in both East-West and North-South  orientations. We first construct a theoretical model of the same star pair with East-West and North-South baseline. We would expect a slower varying oscillation frequency for a North-South baseline compared to an East-West baseline, which is indeed the case as shown in Figure \ref{fig:ew_vs_ns}.

\begin{figure*}
\centering
\includegraphics[width=8.6cm]{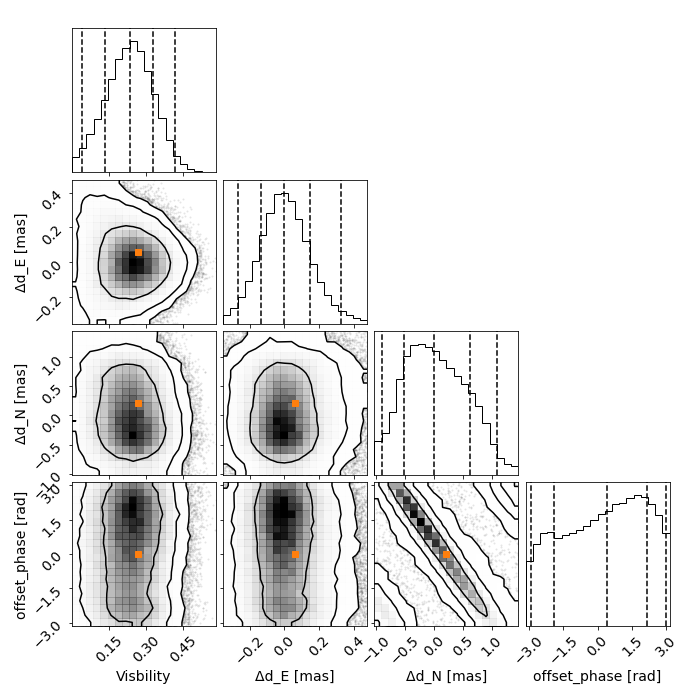}
\includegraphics[width=8.6cm]{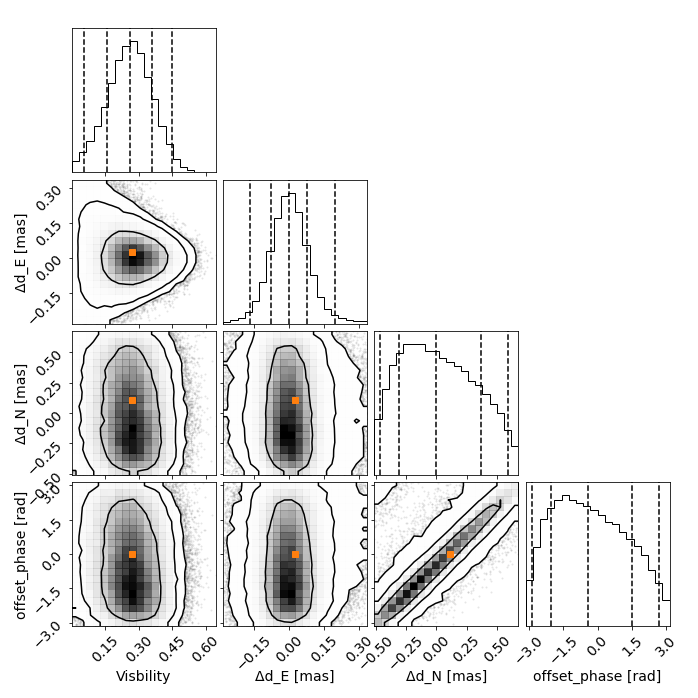}
\includegraphics[width=8.6cm]{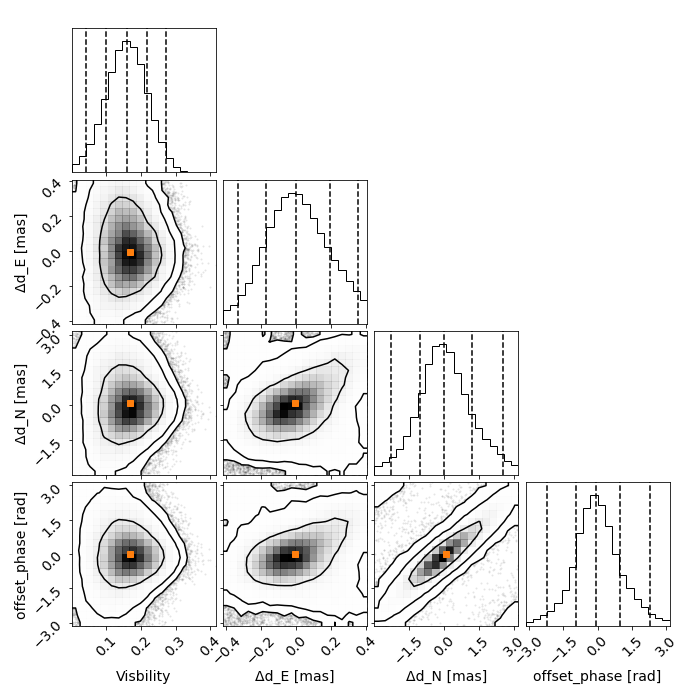}
\includegraphics[width=8.6cm]{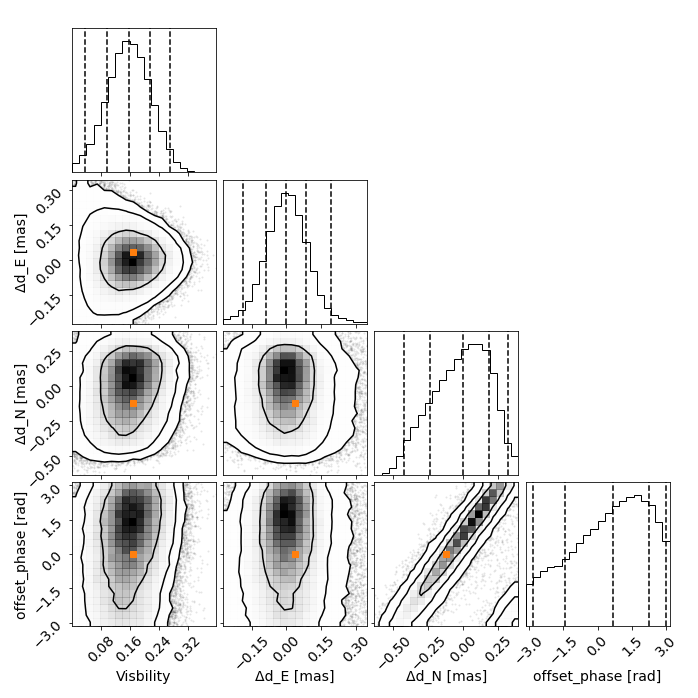}
\caption{These triangle correlation plots are generated using the \texttt{corner} package \cite{corner}. \emph{Top left}: 200 m North-South baseline with a telescope effective collecting area of 1.4 $m^2$ and star pair of 1 arcsec separation. \emph{Top right}: 200 m East-West baseline with a effective collecting area of 1.4 $m^2$ and star pair of 1 arcsec separation. \emph{Bottom left}: 200 m North-South baseline with a telescope effective collecting area of 3.9 $m^2$ and star pair of 15 arcsec separation. \emph{Bottom right}: 200 m East-West baseline with a telescope effective collecting area of 3.9 $m^2$ and star pair of 15 arcsec separation. The 4 parameters used for the MCMC simulation are visibility, $\Delta d_{E}$, $\Delta d_{N}$, and a constant offset phase, $\psi$. The vertical dashed lines represent 2.3 \%, 16 \%, 50 \%, 84 \%, and 99.4 \% quantiles of the Gaussian. $\Delta d_{E}$ and $\Delta d_{N}$ are defined to be the offset from the 50 \% quantile of the Gaussian measured in miliarcseconds. The orange points indicate the true value of each parameter. The contour plots characterizing the correlation between visibility and the two separation parameters have an overall spread behavior as visibility approaches 0. This is expected since likelihood approaches constant as visibility goes to 0. }
\label{fig:triangle_plots}
\end{figure*}

After running the MCMC simulation, the minimum collecting areas needed for obtaining sufficient amount of data to generate well-constrained triangle correlation plots are about 1.4\,m$^2$ for the star pair with 1 arcsec separation and 3.9\,m$^2$ for the pair with 15 arcsec separation. These results are shown in Figure \ref{fig:triangle_plots}. Separation in the East-West and North-South direction are redefined to be the offset from the 50 \% quantile of the Gaussian, i.e. $d_{E} \rightarrow \Delta d_{E}$ and $d_{N} \rightarrow \Delta d_{N}$. As expected, as visibility approaches zero, the contours for the two separation parameters, $\Delta d_{E}$ and $\Delta d_{N}$ broaden, indicating that we are unable to constrain them in that case.

For the 1 arcsec separation case in the top two panels in Figure \ref{fig:triangle_plots} there are no clear correlations between different parameters except for the offset phase and $\Delta d_N$. The configuration with the East-West baseline shows correlation while the North-South baseline shows anti-correlation. This is expected as explained below. Since the 1 arcsec pair has the same right ascension, i.e. $d_N = 0$, their $\Delta L$ in a pure North-South and East-West baseline configuration becomes $\Delta L_{ns} = B_Nd_N(\sin{\theta_L}\sin{\delta_{mid}}\cos{\phi_{mid}} +\cos{\theta_L} \cos{\delta_{mid}} )$ and $\Delta L_{ew} = -B_Ed_N\sin{\phi_{mid}}\sin{\delta_{mid}} $, respectively. If we were to remove the constant term $B_Nd_N\cos{\theta_L} \cos{\delta_{mid}}$ for $\Delta L_{ns}$, it has been tested that we would get back a correlating relation between $d_N$ and the constant offset phase, $\psi$. 

For the 15 arcsec separation case in the bottom two panels in Figure \ref{fig:triangle_plots}, $d_E$ appears to be correlating with both offset phase and $d_N$ for the North-South baseline case while showing no correlation for the East-West baseline. These correlations should go away once there is enough statistics, and one easy way is to simply increase the effective collecting area of the telescopes. 

Table \ref{tab:par_error} summarizes the errors for different parameters of different baseline configurations and effective collecting area for the considered two star pairs. As the collecting area of the telescope decreases the overall trend of error increase of different parameters is observed as expected. The offset phase does not seem to be well constrained compared to other parameters except for the 15 arcsec star pair with a North-South baseline configuration. $\sigma_{\psi}$ for 15 arcsec pair with North-South configuration seems to be a lot smaller compared to other cases while $\sigma_{N}$ becomes really large.

\begin{table*}[ht]

\caption{\label{tab:par_error}%
Summary of uncertainties for different parameters derived from the MCMC simulation. The second column shows the separation of the star pair in arcsec and the collecting area of the telescope in square meters. The third column shows the visibility of each star pair. The next columns show the standard error for visibility ($\sigma_V$), separation in the East-West ($\sigma_E$) and north-South direction ($\sigma_N$) in mili-arcseconds, and the constant phase ($\sigma_\psi$) in radians. 
}

\begin{ruledtabular}
\begin{tabular}{cccccccc}
\textrm{Baseline Configuration} &
\textrm{Distance [arcsec] /Area [$m^2$]} & 
\textrm{Visibility} &
\textrm{$\sigma_V$} & 
\textrm{$V/\sigma_V$}&
\textrm{$\sigma_{E}$ [mas]} &
\textrm{$\sigma_{N} $ [mas]}& 
\textrm{$\sigma_{\psi}$ [rad]}\\ 
\colrule
East-West & 15, 3.9 &  0.1692 & 0.0591 & 2.863 & 0.0910 & 0.1977 & 1.6401 \\

East-West & 15, 4.2 &  0.1692 & 0.0530 &  3.192 & 0.0863 & 0.1941 & 1.6170 \\

East-West & 15, 5.0 &  0.1692 & 0.0446 & 3.794 & 0.0807 & 0.1925 & 1.6301 \\

East-West & 1, 1.4 & 0.2683  & 0.0999 & 2.686 & 0.0845 & 0.3300 &  1.6434 \\

East-West & 1, 1.8 & 0.2683 & 0.0781 & 3.435 & 0.0668 & 0.3002 & 1.6184 \\

East-West & 1, 2.1 & 0.2683 & 0.0656 & 4.090 & 0.0511 &  0.2910 & 1.5536 \\

North-South & 15, 3.9 & 0.1692 & 0.0584 & 2.897 & 0.1751 & 1.1553 & 1.0484\\

North-South & 15, 4.2 & 0.1692 & 0.0566 & 2.989 & 0.1327 & 1.4287 & 0.9458 \\ 

North-South & 15, 5.0 & 0.1692 & 0.0480 & 3.525 & 0.1147 & 1.1665 & 0.8035 \\

North-South & 1, 1.4 & 0.2683 & 0.0945 & 2.839 & 0.1433 & 0.5258 & 1.7407 \\

North-South & 1, 1.8 & 0.2683 & 0.0764 & 3.512 & 0.1022 & 0.4240 & 1.5434\\

North-South & 1, 2.1 & 0.2683 & 0.0651 & 4.121 & 0.0966 & 0.4424 & 1.6060 \\

\end{tabular}
\end{ruledtabular}
\end{table*}


\section{Summary and Outlook}
\label{sec:conclusions}

We have investigated and verified via direct simulations observability of useful signals in the novel two-photon interferometer proposed in~\cite{Stankus_2022} to use for high-precision astrometry.  Here we have expanded beyond the general estimates presented in~\cite{Stankus_2022} to model the photon arrival data from several real pairs of stars, which would be appropriate targets for a demonstration experiment. Our results are consistent with the previous estimates made using a Fisher matrix calculation, and we have identified the effective size of the telescope collecting aperture that will lead to a significant detection in a single night.

We recognize that it is non-trivial to establish two astronomy-grade light collectors each with effective collecting area of 3.9~m$^2$, made even harder since effective area will be smaller than geometric area due to various losses, e.g. fiber coupling, detector quantum efficiency, etc.  However, we also note that the calculation here assumes only a very narrow optical band is being used, just $\Delta \nu \sim$~1~GHz as in Section~\ref{sec:selection} above.  Plenty more photons are available, and with spectrographic separation onto an array of detectors, as suggested in Figure~\ref{fig:idea}, we can effectively run many experiments in parallel using the same collecting apertures.  Using $N$ instrumented spectrographic channels would allow us to collect $N$ times as many pairs, or equivalently to reduce the collecting area by a factor of $1/\sqrt{N}$.  So, for example, instrumenting 100 spectral channels of $\Delta\nu \sim$1~GHz each -- still a tiny fraction of the optical band in total -- would yield the same precision as described above but with only 0.39~m$^2$ of collecting area for each telescope.

We have written down the exact Bayesian analysis framework for the corresponding data analysis. This enables one to calculate the posterior probability for any model with an appropriate summation over the event timestamps. The posterior probability can then be explored using any of the standard techniques, which in our case we relied on the standard MCMC approach. This should be sufficient for the foreseeable future. The source code \texttt{QA-sim} used for the simulations can be found in the Github page \cite{QA-sim}.

As the number of observed coincidences increases, the exact likelihood will have to be eventually replaced with approximate methods that will likely involve a Fourier transformation. We leave this exploration for the future work.

It was not in the scope of this work to consider all possible systematic effects that can limit the achievable resolution as the primary goal was to determine the hard limit coming from the photon statistics. Nevertheless in regard to the obvious topic of atmospheric fluctuations, we note that techniques such as adaptive optics will be applicable here as well \cite{meyer2010,cameron2008,davies2012}. We leave these investigations to the future experimental work.


Overall we consider the proposed approach as something that is technically feasible and could be experimentally tested with existing technologies of single photon detection in the near future \cite{Nomerotski2020_1}.

\section*{Acknowledgements}
This work was supported by the U.S. Department of Energy QuantISED award and BNL LDRD grant 19-30. Zhi Chen acknowledges support under the Science Undergraduate Laboratory Internships (SULI) Program  by the U.S. Department of Energy. 

We also acknowledge other Python software such as \texttt{Matplotlib} \cite{matplotlib}, \texttt{Numpy} \cite{numpy}, and \texttt{Scipy} \cite{scipy} for the extensive usage during the development of this project.

\vspace*{1cm}

\bibliography{references.bib}

    

\end{document}